\documentclass{nature_modifbyHA}
\usepackage{amsmath}
\usepackage{graphicx}
\newcommand{\V}[1]{\textup{\scriptsize{#1}}}

\title{Superconducting parity effect across the Anderson limit} 

\author
{Sergio Vlaic,$^{1}$ St\'ephane Pons,$^{1}$ Tianzhen Zhang,$^{1}$ Alexandre Assouline,$^{1}$\\ Alexandre Zimmers,$^{1}$ Christophe David,$^{2}$ Guillemin Rodary,$^{2}$\\ Jean-Christophe Girard,$^{2}$ Dimitri Roditchev,$^{1}$ Herv\'e Aubin$^{1}$\\
}

\begin{document} 

\maketitle
\begin{affiliations}
\item{LPEM, ESPCI Paris, PSL Research University; CNRS; Sorbonne Universit\'es, UPMC University of Paris 6,10 rue Vauquelin, F-75005 Paris, France}
\item{Centre de Nanosciences et de Nanotechnologies, CNRS, Univ. Paris-Sud, Universit\'es Paris-Saclay, C2N – Marcoussis, 91460 Marcoussis, France}
\end{affiliations}

\begin{abstract}

How small superconductors can be? For isolated nanoparticles subject to quantum size effects, P.W. Anderson conjectured in 1959 that superconductivity could only exist when the electronic level spacing $\delta$ is smaller than the superconducting gap energy $\Delta$.

Here, we report a scanning tunneling spectroscopy study of superconducting lead (Pb) nanocrystals grown on the (110) surface of InAs. We find that for nanocrystals of lateral size smaller than the Fermi wavelength of the 2D electron gas at the surface of InAs, the electronic transmission of the interface is weak; this leads to Coulomb blockade and enables the extraction of the electron addition energy of the nanocrystals. For large nanocrystals, the addition energy displays superconducting parity effect, a direct consequence of Cooper pairing. Studying this parity effect as function of nanocrystal  volume, we find the suppression of Cooper pairing when the mean electronic level spacing overcomes the superconducting gap energy, thus demonstrating unambiguously the validity of the Anderson criterion.

\end{abstract}

The addition energy of an electron to a superconducting island, weakly coupled to the environment by the capacitance $C_{\small{\Sigma}}$, is given by (See Methods):

\begin{equation}
E_{\V{even (odd)}}=\frac{e^2}{C_{\small{\Sigma}}}+(-)2\Delta+\delta
\end{equation}
	
where the first term is the Coulomb energy, the second term depends on the parity of electron occupation number as a consequence of the formation of a Cooper pair\cite{Averin1992,Lafarge1993b}, the third term is the electronic level spacing in the island. This parity effect has been observed in large $\sim 1~\mu$m micro-fabricated Al islands, through direct measurement of the charge capacitance of the island\cite{Lafarge1993b}, through the even-odd modulation of the addition energy in single electron transistors\cite{Tuominen1992,Eiles1993,Lafarge1993a,Higginbotham2015} or the parity dependence of the Josephson current in Cooper pair transistors\cite{Joyez1994,Aumentado2003,VanWoerkom2015}.

Until now, the parity effect on the addition energy has never been observed in small nanocrystals (NCs) near the Anderson limit~\cite{Anderson1959}, reached at a volume about $V_{\V{Anderson}}\simeq$ 100 nm$^3$, where the mean electronic level spacing $<\delta>$ equals the superconducting gap energy $\Delta$.

In single electron transistors fabricated with nanosized superconducting grains of aluminum\cite{Ralph1995,VonDelft2001}, the 2e modulation of the addition energy could not be observed directly. Also, because only a few devices could be fabricated, testing the Anderson criterion was not possible with this approach. Indirect indications for the disappearance of superconductivity in small superconducting grains came from magnetization measurements \cite{Reich2003,Zolotavin2010}; because these measurements were averaged over macroscopic quantities of NCs, the link to the Anderson limit remained ambiguous. In this work, we present a new system that enables a study of single and isolated NCs across the Anderson limit, where the NCs can be reproducibly obtained in large quantities. The superconducting gap energy and the transition temperature are measured through a study of the superconducting parity effect in the addition energy of the NCs. This constitutes an alternative approach to conventional tunneling measurement of the superconducting gap in the quasi-particle excitation spectrum, which cannot be accessed as a consequence of the Coulomb gap at zero bias.

%instead, the occupation parity was determined by studying the evolution of the discrete spectrum in an applied magnetic field\cite{Ralph1995,Black1996,Ralph1997}, or by looking for the presence of the superconducting gap in the tunneling spectrum\cite{Black1996,Ralph1997}.

\section*{Results}
\subsection{Sample preparation}
The observation of the parity effect is challenging as it requires clean systems, free of impurity states responsible for the so-called quasiparticle poisoning\cite{Savin2007,VanWoerkom2015}. Furthermore, scanning tunneling spectroscopy of isolated NCs requires, in addition to the tip-NC tunnel barrier, a second tunnel barrier between the NC and the conducting substrate\cite{Hong2013a,Bose2010}, as sketched in Supplementary Fig. 1.

In this work, the Pb NCs are obtained by thermal evaporation of a nominal 0.3 monolayer of Pb on the (110) surface of InAs heated at $T=150~^\circ$C. The (110) surface is obtained by cleaving a n-type InAs substrate in ultra high vacuum at a base pressure $P\sim 10^{-10}$~mbar.  Two distinct samples (A and B) have been prepared with slightly different NC concentrations and sizes. The volume of the NCs ranges from 20 nm$^3\simeq 0.2~$ $V_{\V{Anderson}}$ to 800 nm$^3\simeq 8~V_{\V{Anderson}}$ while the height ranges from 1 unit cell (0.495 nm) to 5.2 nm, see Supplementary Fig. 2 and Supplementary Note 1 for details on NC volume determination. The scanning tunneling microscopy (STM) topographic images, Fig.~1abc and  Supplementary Fig. 3, for sample A and sample B, respectively, show that Pb grows in the Volmer-Weber, i.e. island mode\cite{Brune1998}. The 3D Laplacian image $\Delta_{xy}z(x,y)$, Fig.~1c, shows that the NCs are well crystallized and expose mostly the (111) planes of the cubic face-centered Pb structure, as indicated by the observation of the characteristic hexagonal shape of the (111) facets. Surrounding these NCs, the surface remains free from adsorbate, as atomic resolution images of the (110) InAs surface prove, Fig.~1d.

\subsection{Tip-induced QDot on the InAs surface}
Fig.~1e shows the Differential Conductance (DC) $dI/dV$ measured on the InAs surface at several distances, from 0 to 10 nm, of a Pb NC. The data are measured at $T= 1.3~$K, unless indicated otherwise, using a standard lock-in procedure (See Methods). The data indicate that the Fermi level is in the conduction band of InAs as expected for this n-doped sample. With a sulphur dopant concentration, $N_\V{D}\sim 6\times10^{16}$~cm$^{-3}$, the Fermi level is 21 meV above the conduction-band minimum. A zoom on these spectra, Fig.~1f, shows multiple peaks that result from the discrete levels of the tip-induced Quantum Dot (QDot), a phenomena that has also been observed in previous works\cite{Dombrowski1999}. This demonstrates that Pb deposition on InAs do not produce any significant defects and doping. Indeed, in presence of defects or adsorbate, the surface of III-V semiconductors present interface states that pin the Fermi level at the charge neutrality level\cite{Tersoff1984a,Monch2001}, Fig.~1g. For InAs, this level is located 150 meV above its conduction band minimum, which leads to the formation of an electron accumulation layer as shown by numerous photoemission experiments\cite{Tersoff1984a,Morgenstern2012a}. In contrast, perfectly clean (110) surfaces do not present any interface states and consequently the Fermi level is not pinned. Thus, the electric field from the STM tip can easily shift the conduction band and generates the so-called tip-induced QDot\cite{Dombrowski1999}, as sketched Fig.~1h. While the energy of the QDot levels can shift on long distances, see Supplementary Fig. 4, as a consequence of variations in the electrostatic environment due to the random distribution of Pb NCs and sulphur dopants, we see, Fig.~1f, that the QDdot levels are not altered on short distances ($<$ 10 nm) near the NCs. Only a weak broadening of the QDot levels is observed, likely a consequence of their weak tunnel coupling with the Pb NCs.

\subsection{Coulomb blockade and nature of the tunnel barrier}
On NCs of three distinct sizes shown Fig.~2abc, representative DC spectra are shown Fig.~2de. They display a Coulomb gap at zero bias of width $\delta V_{\V{sub}}=e/(C_{\V{sub}}+C_{\V{tip}})$  where $C_{\V{sub}}$ ($C_{\V{tip}}$) is the capacitance between the NC and the substrate (tip). The data also display sharp Coulomb peaks where the voltage interval between the peaks provides the addition voltage $\delta V_{\V{add}}$ for an electron, which is related to the addition energy by : $\delta V_{\V{add}}=E_{\V{add}}/e\eta$ where $\eta=\frac{C_{\V{tip}}}{C_{\V{tip}}+C_{\V{sub}}}$ is the arm lever; see Methods section for a derivation of these relations. Furthermore, the DCs may also display broad additional peaks, of weak amplitude in large NCs, $V/V_\V{Anderson}>1$, as indicated by arrows in Fig.~2d, but of large amplitude in small NCs, $V/V_\V{Anderson}<<1$, as indicated by arrows in Fig.~3. These broad peaks are the signature of quantum well states in the Pb NCs due to strong confinement in the  $<111>$ direction as observed in scanning tunneling studies of thin layers of Pb\cite{Su2001}.

The color map Fig.~2e shows that $\delta V_{\V{add}}$ changes slightly with the tip position above the NC, as consequence of the variation in the tip-NC capacitance $C_{\V{tip}}$. Fig.~3 shows the DCs for 13 additional NCs, from which the capacitance $C_{\V{sub}}$ is extracted and shown as colored symbols in Fig.~2f and Supplementary Fig. 5. On these last plots, data points shown as black circles of 24 other NCs are also included, for which the DCs are not shown. Fig.~2f shows that $C_{\V{sub}}$ increases linearly with the area $A$ as $C_{\V{sub}}=A \varepsilon/d$, using $\varepsilon$ = 12.3, the dielectric constant of InAs and $d$ = 4~nm for the effective tunnel barrier thickness. 

As no dielectric insulator has been deposited on the surface and no Schottky barrier exists at metal-InAs interfaces\cite{Monch2001,Morgenstern2012a}, the origin of the tunnel barrier and the meaning of the thickness $d$ appear clearly only after one realizes that the Fermi wavelength of the 2D gas in InAs is larger than the lateral size of the NCs. At the interface between the Pb NC and InAs, the Fermi energy in InAs is at the charge neutrality level, $E_{\V{F}}=150$~meV, Ref.~\cite{Monch2001,Morgenstern2012a}, which gives for the Fermi wavelength $\lambda_{\V{F}}=20$~nm. As known from numerous works with quantum point-contacts formed in 2D electron gas\cite{Contacts1988,Pasquier1993}, the transmission coefficient $T$ decreases for constrictions smaller than the Fermi wavelength. Because a NC covers only a fraction of the area $\simeq\lambda_{\V{F}}^2$, its transmission coefficient with the 2D gas is significantly smaller than one, which explains the observation of the Coulomb blockade. For a small NC, the weak coupling model Ref.~\cite{Hanna1991} can be used to describe the data, as shown Fig.~2g. This model shows that the contact impedance is of the order of $R_{\V{contact}}\sim 10$~M$\Omega$, implying that the transmission coefficient $T=R_{\V{contact}} e^2/h=0.0025$ is weak as anticipated. In this model, the magnitude of the Coulomb peaks increases with the ratio $R_{\V{tunnel}}/R_{\V{contact}}$, as observed on the DC curves measured as function of tip height, Supplementary Fig. 6. Fig.~2h shows the amplitude of the Coulomb peak, normalized to its base value, as function of NC area. The amplitude is constant for small area ($<$ 100 nm$^2$) but decreases quickly for area approaching $\pi\lambda_{\V{F}}^2/4\simeq300$~nm$^2$. This behavior cannot be described by the weak-coupling model just discussed, however, it can be understood by considering models of Coulomb blockade in the strong coupling regime\cite{Matveev1995,Aleiner2002}. These models show that the Coulomb oscillations disappear when $T$ approaches unity, when charge fluctuations between the NC and the substrate become significant. Fig.~3 shows that the Coulomb peaks of the largest NCs have almost completely disappeared. The fact that the amplitude of the Coulomb peaks decreases for NCs area approaching $\lambda_{\V{F}}^2$ confirms our interpretation that the tunnel barrier is due to a quantum constriction of the electronic wave function at the interface between the NC and the 2D gas. Thus, the dielectric thickness $d$ = 4 nm extracted from $C_{\V{sub}}$ above is actually set by the Debye length of the 2D gas and $C_{\V{sub}}$ actually corresponds to the quantum capacitance of InAs.

\subsection{Superconducting parity effect}
Thanks to this highly clean type of tunnel junction, free from quasi-particle poisoning, the superconducting parity effect in the NCs can be observed through the even-odd modulation of the addition voltage, as shown Fig.~2de, Fig.~3 and Fig.~4. The addition voltages can be precisely extracted thanks to the sharpness of the Coulomb peaks, which voltage positions are obtained through a fit with a Lorentz function, Supplementary Fig. 7. As sketched Fig.~2i and shown by Eq.~1, the addition voltage $\delta V_{\V{even}}$ for injecting an electron in an even parity NC is higher than $\delta V_{\V{odd}}$ for injecting an electron in an odd parity NC, where the energy difference is given by the binding energy of the Cooper pair. Fig.~4a shows the DCs for a large NC, $V/V_\V{Anderson}$ = 1.6, as function of temperature. The corresponding addition voltages, shown Fig.~4b, are almost equal above $T_\V{c}=7.2$~K, the superconducting transition temperature of bulk Pb. However, an even-odd modulation is observed at low temperature $T=1.3$~K. The difference in the addition energies between two successive charge configurations is obtained from $\delta E =e\eta(\delta V_{\V{even}}-\delta V_{\V{odd}})$. For this large NC, four Coulomb peaks are observed which provide three distinct addition voltages indicated by the horizontal bars. From these addition voltages, two distinct values of the addition energy difference $\delta E$ between two charge configurations are obtained and given by  $\delta E=\eta(\delta V_{\V{Head}}-\delta V_{\V{Tail}})$, where the head (tail) refers to the colored arrows in the panel. These two values of $\delta E$ are shown Fig.~4c as function of temperature. Their values is near zero at high temperature, $\delta E_{\V{HT}}\sim 0$, and increase below $T_{\V{c}}=7.2~$K to reach, at low temperature, the theoretically expected value $|\delta E_{\V{LT}}|\sim 4\Delta_{\V{bulk}}$, Ref.~\cite{Averin1992}, where $\Delta_{\V{bulk}}=1.29$~meV is the superconducting gap of bulk Pb. The value $\delta E_{\V{LT}}$ changes sign as one goes from the difference between two addition energies $\delta E =e\eta(\delta V_{\V{even}}-\delta V_{\V{odd}})$ to the next difference $\delta E =e\eta(\delta V_{\V{odd}}-\delta V_{\V{even}})$.

For NCs smaller than the Anderson volume, Fig.~4dgj, we observe that $\delta E_{\V{HT}}$ is non-zero, which indicates that the electronic level spacing $\delta$ has now a significant contribution to the addition energy, following Eq.~1. The values of $\delta E_{\V{HT}}$ are distinct between successive charge configurations. Indeed, in metallic systems, the electronic levels are randomly distributed as described by  Random Matrix Theory (RMT)\cite{Alhassid2000}. Collecting the values $\delta E_{\V{HT}}$ for all NCs, Fig.~5a shows that, in average, the evolution of $\delta E_{\V{HT}}$ with NC volume can be properly described by the relation:

\begin{equation}
<\delta>=\frac{2(\pi\hbar)^2}{m^* k_{\V{F1(2)}} \textup{Volume}}
\end{equation}

using m$^*$=1.2 m$_e$ for the effective mass, where $k_{\V{F1}}$=7.01 nm$^{-1}$ and $k_{\V{F2}}$=11.21 nm$^{-1}$ are characteristic wave-vectors of the two Fermi surfaces FS1 and FS2 of Pb.

For the NC of volume $V/V_\V{Anderson}$ = 0.89, Fig.~4d, while the level spacing $\delta E_{\V{HT}}$ is large, the shift of the Coulomb peaks due to the parity effect is still dominating the temperature dependence and can be observed directly on the raw data and the addition energy difference $\delta E$ plotted as function of temperature on Fig.~4e. A line $\delta E_{\V{HT}_{\V{ext}}}$ is extrapolated from high temperature and the difference  $\delta E(T) - \delta E_{\V{HT}_{\V{ext}}}(T)$ gives the temperature dependence of the superconducting gap, Fig.~4f, which shows that the critical temperature $T_\V{c} \simeq 6~$K is smaller than the bulk value. The amplitude of the superconducting gap is obtained from $\Delta=(\delta E(T=1.2 K) - \delta E_{\V{HT}_{\V{ext}}}(T=1.2~ \rm K))/4$. For this NC, the superconducting energy gap is about two times smaller than the bulk value, $\Delta=\Delta_{\V{bulk}}/2$.
 
For the smaller NC of volume $V/V_\V{Anderson}$ = 0.55, Fig.~4g, the level spacing $\delta E_{\V{HT}}$ is larger and has a temperature dependence that dominates the shift of the Coulomb peaks with temperature. This shift could be the consequence of thermally induced electro-chemical shifts or temperature dependent strain or electric field effects. While the parity effect is barely visible on the raw data, using the procedure employed for the previous NC, the temperature $T_\V{c} \simeq 5$~K value and the energy gap $\Delta\simeq\Delta_{\V{bulk}}/4$ can be extracted, Fig.~4i.

Finally, for the smallest NCs $V/V_\V{Anderson}$ = 0.43 (0.34), shown respectively Fig.~4j and Supplementary Fig. 8h, they have the largest level spacing $\delta E_{\V{HT}}$ and, even though the addition energies are measured with much higher resolution than the superconducting gap energy, no parity effect can be observed on Fig.~4l and Supplementary Fig. 8j, respectively.

For 13 NCs where the DCs have been acquired as function of temperature, some of which are shown Fig.~4 and Supplementary Fig. 8, the level spacing, the superconducting gap energy and the transition temperature are extracted and plotted Fig.~5a, Fig.~5b and Fig.~5c, respectively. Upon reducing the NC volume, both quantities display a sharp decrease to zero when the level spacing becomes of the order of the superconducting gap energy, $\simeq$ 1 meV. See Supplementary Note 2 for a comparison with the results of Bose et al.\cite{Bose2010} on a system where the superconducting nanoparticles are strongly coupled to the normal substrate.

\section*{Discussion}
Fig.~5 suggests that superconductivity disappears when the mean level spacing at the Fermi surface of the electron-type band, Fig.~5e, increases up to the superconducting gap energy. This is consistent with recent theoretical calculations \cite{Floris2007} and STM measurements\cite{Ruby2015} which have shown that electron-phonon coupling is stronger for this electron-type band owing to its p-d character. Regarding the BCS ratio, within the experimental resolution, no significant deviation from the bulk value has been observed.

To summarize, we have found that a 2D electron gas of large Fermi wavelength constitutes an ideal substrate for studying Coulomb blockade in nanosized NCs evaporated in an ultra-high vacuum environment. This discovery leads us to observe, for the first time by STM, the parity effect and quantum confinement in isolated superconducting NCs and enabled the first demonstration of the Anderson criterion for the existence of superconductivity at single NC level. Furthermore, this new insight on the superconductor-InAs interface is of interest for topological superconductivity where Majorana islands are generated by depositing a superconductor on InAs nanowires\cite{Das2012,Albrecht2015}.

\begin{methods}
\subsection{Relation between sample bias and energies}
The Coulomb gap at zero bias results from Coulomb blockade that prevent charge fluctuations in the NC. As sketched in Supplementary Fig. 1, Coulomb blockade is lifted when the Fermi level of either one of the electrodes is aligned with one of the excited levels of the NC. Thus the amplitude of the Coulomb gap observed in the DC is given by $\delta V_{\V{sub}}=\frac{e}{C_{\small{\Sigma}}}$=2$\times\frac{ E_{\V{C}}}{e}$, with $E_{\V{C}}=\frac{e^2}{2C_{\small{\Sigma}}}$.

The Coulomb peaks observed at higher voltages result from the shift of the electrochemical potential of the NC upon increasing the voltage bias across the double junction. This shift is given by :
\begin{eqnarray}
\label{Eq1}
\frac{\Delta \mu}{e}=\eta V_{\rm Bias}
\end{eqnarray}
with :
\begin{eqnarray}
\label{Eq2}
\eta=\frac{C_{\V{tip}}}{C_{\V{tip}}+C_{\V{sub}}}
\end{eqnarray}

Charge states with increased number of electrons become accessible when the electrochemical potential changes by $2\times E_{\V{C}}$. Thus the voltage difference between two charge states is given by :

\begin{eqnarray}
\label{Eq1}
\Delta V_{\rm add}=\frac{1}{\eta} \times \frac{2E_{\V{C}}}{e}=\frac{e}{C_{\V{tip}}}
\end{eqnarray}

This formula shows that the addition voltage depends only on the capacitance C$_{\V{tip}}$ and not on the capacitance C$_{\V{sub}}$, as shown Supplementary Fig. 1b, where a simulation of the conduction spectrum, using the Hanna and Tinkham model\cite{Hanna1991} for two distinct values of the capacitance C$_{\V{sub}}$.

\subsection{Addition energies}

Following Ref.~\cite{Kouwenhoven2001,Averin1992}, the total energy of a NC with N electrons is given by:

\begin{align}
E(N)=\frac{(N e)^2}{2C_{\small{\Sigma}}}+E_0(N)\nonumber\\
\\
E_0(N)=
\begin{cases}
\Delta \text{ for odd N,}\nonumber\\
\text{ 0 for even N,}
\end{cases}
\end{align}

The electrochemical potential of a nanocrystal with an even (odd) N (N+1) number of electrons is given by:
\begin{align}
\mu(N)=E(N+1)-E(N)=(N+\frac{1}{2})\frac{e^2}{C_{\small{\Sigma}}}+\Delta\nonumber\\
\\
\mu(N+1)=E(N+2)-E(N+1)=(N+\frac{3}{2})\frac{e^2}{C_{\small{\Sigma}}}-\Delta\nonumber\\
\nonumber
\end{align}

From these last equations, one obtains the addition energies for a NC with an even (odd) N (N+1) number of electrons :

\begin{align}
E_{\V{even}}=\mu(N)-\mu(N-1)=\frac{e^2}{C_{\small{\Sigma}}}+2\Delta\nonumber\\
\\
E_{\V{odd}}=\mu(N+1)-\mu(N)=\frac{e^2}{C_{\small{\Sigma}}}-2\Delta\nonumber\\
\nonumber
\end{align}

Thus, the difference of addition energies between two successive charge states is given by :

\begin{align}
\delta E= E_{\V{even}}- E_{\V{odd}}=4\Delta\\
\end{align}

When the electronic spectrum of the NC is discrete, the level spacing $\delta$ should be included in the addition energy.
\begin{align}
E_{\V{even (odd)}}=\frac{e^2}{C_{\small{\Sigma}}}+(-)2\Delta+\delta\\
\end{align}

\subsection{Random level distribution}
In metallic NCs, the electronic level distribution is described by RMT\cite{Alhassid2000,Halperin1986}. In a NC with strong spin-orbit coupling, RMT predicts that the level spacing should be described by a Gaussian symplectic ensemble. For this level distribution, shown Supplementary Fig. 9, the width of the distribution, i.e. the standard deviation, is equal to $\sigma \simeq <\delta>$,\cite{Alhassid2000,Mehta2004}. Between two successive charge states, the addition energy can fluctuate by an amount of the order of $\sigma$, consequently, in average, the difference in addition energies between two successive charge states is given by:
\begin{align}
\delta E_{\V{LT}} = E_{\V{even}}- E_{\V{odd}}=4\Delta+<\delta>\\
\end{align}

At temperatures above the superconducting transition temperature:
\begin{align}
\delta E_{\V{HT}}=<\delta>\\
\end{align}

Thus an estimation of the level spacing can be obtained by a measure of the difference in the addition energies above $T_{\V{C}}$.

Furthermore, the gap amplitude can be obtained from:
\begin{align}
\Delta=(\delta E_{\V{LT}}-\delta E_{\V{HT}})/4
\end{align}

\subsection{Measurements details}

The microscope used is a low temperature, $T_{\V{base}}=1.3$~K, Joule-Thomson JT-STM from SPECS accommodated with a preparation chamber operating in Ultra High Vacuum at a base pressure $P\sim 10^{-10}$~mbar.
The differential conductance curves dI/dV are measured with a standard lock-in procedure. An AC signal of amplitude about $\simeq$ 1~meV and frequency about 777~Hz is employed.

\subsection{Data avaibility}
The data that support the main findings of this study are available from the corresponding author upon request.

\end{methods}

\begin{figure}
	\begin{center}
		\includegraphics[width=16cm]{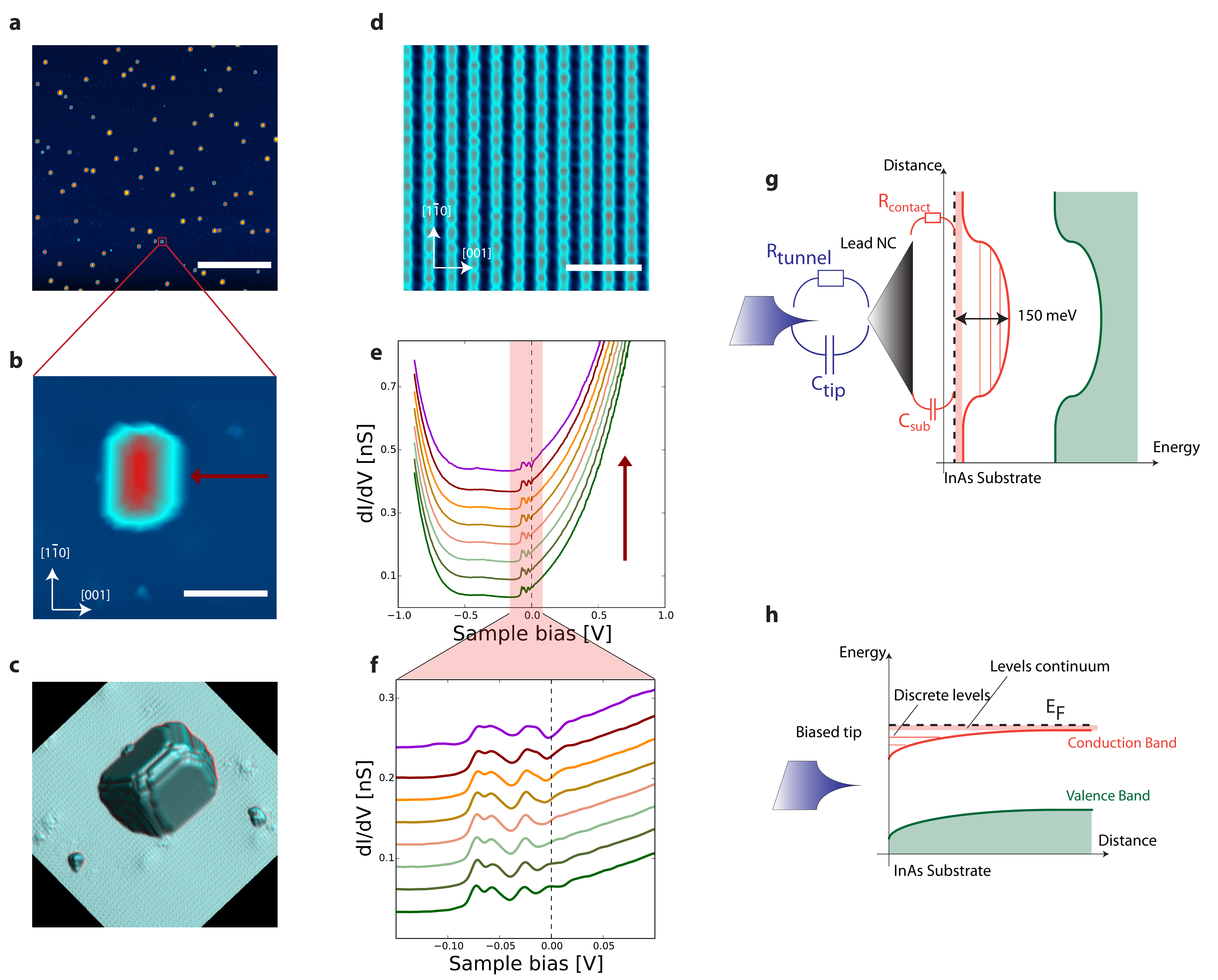}
		\caption{\label{Fig1} \textbf{Pb NCs on InAs (110).} (\textbf{a}) 1~$\mu$m$\times$1~$\mu$m topographic STM image (1 V, 30 pA) of Pb NCs grown on the (110) InAs surface of sample A. Scale bar is 300 nm. (\textbf{b}) Zoom on 30 nm$\times$30 nm area, showing a Pb NC. Scale bar is 10 nm. (\textbf{c}) 3D Laplacian  $\Delta_{xy}z(x,y)$ image of a NC. (\textbf{d}) 6.5~nm$\times$6.5~nm atomic resolution image of InAs (110) obtained near the NC. Scale bar is 2 nm. (\textbf{e}) DC measured at several distances from the  Pb NC along the red arrow on panel \textbf{b}. (\textbf{f}) Zoom at low bias showing the conductance peaks due the discrete levels of the tip-induced quantum dot. (\textbf{g}) Sketch of the band bending below the Pb NC due to the pinning of the Fermi level at the charge neutrality level. (\textbf{h}) Sketch of the band bending induced by the tip leading to the formation of a quantum dot.}
	\end{center}
\end{figure}

\begin{figure}
	\begin{center}
		\includegraphics[width=16cm]{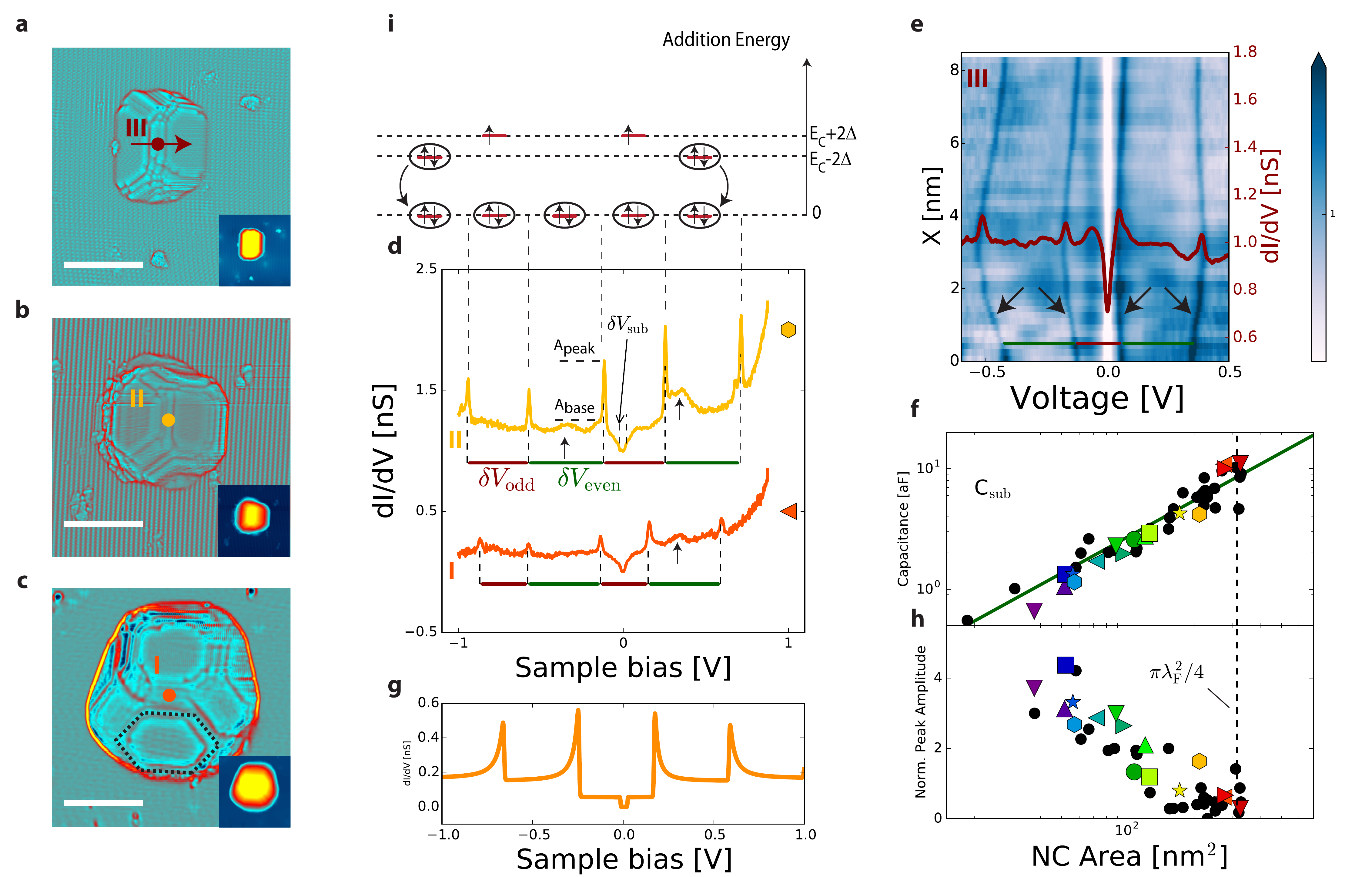}
		\caption{\label{Fig2} \textbf{Pb NCs in the regime of Coulomb blockade.} (\textbf{a},\textbf{b},\textbf{c}) 30 nm $\times$ 30 nm Laplacian $\Delta_{xy}z(x,y)$ topographic images (30 pA, 1V) of NCs of decreasing size, labeled I to III, where the hexagonal shape of the (111) facets is visible, as shown by the dash line on panel \textbf{c}. The scale bars correspond to 10 nm. The insets show the corresponding topographic STM images. (\textbf{d}) DC measured at the center of NC I and II, indicated by dots on the panels \textbf{b},\textbf{c}. The addition voltages $\delta V_{\V{odd}}$ and $\delta V_{\V{even}}$ can be identified for each curve. The black arrows indicate the local maxima in the density of states due to quantum well states.
		The colored symbols identify the corresponding data points in panels \textbf{f,h} and Fig.~5. (\textbf{e}) DC map as function of sample bias and distance measured on NC III along the red arrow shown panel \textbf{a}. The black arrows indicate the Coulomb peak lines. (\textbf{f}) Capacitance C$_{\V{sub}}$ extracted from the Coulomb gap at zero bias. It scales linearly with the NC area. (\textbf{g}) Simulation of the DC for NC II using the weak coupling model\cite{Hanna1991}. (\textbf{h}) Normalized Coulomb peak amplitude $A_{\V{norm}}=(A_{\V{peak}}-A_{\V{base}})/A_{\V{base}}$, this value decreases at the approach of the area $\pi\lambda_{\V{F}}^2/4$. (\textbf{i}) Sketch of electron occupation of NC II.}
	\end{center}
\end{figure}

\begin{figure}
	\begin{center}
		\includegraphics[width=10cm]{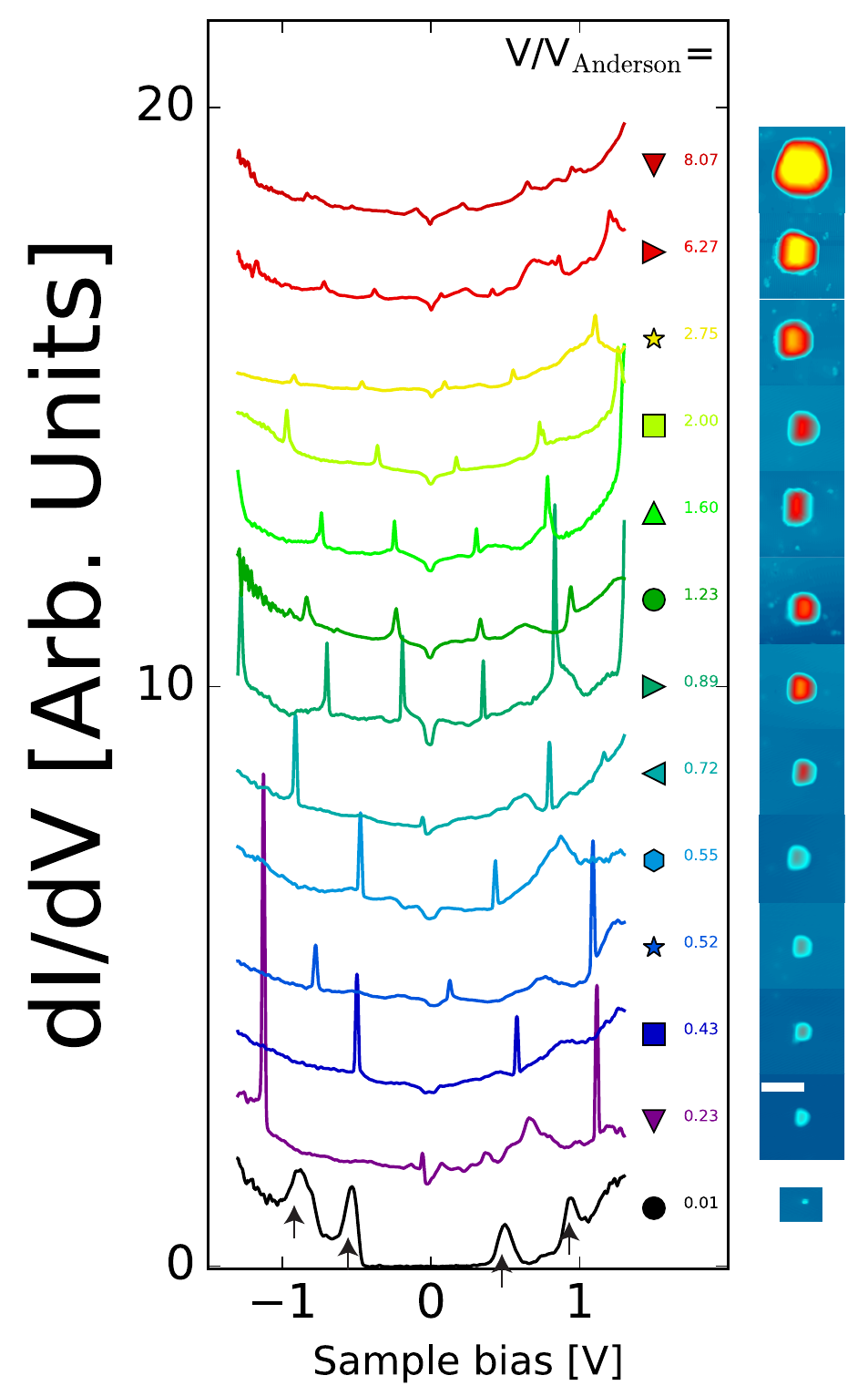}
		\caption{\label{Fig3} \textbf{DCs for increasing NC volume.}  The colored symbols identify the corresponding data points in Fig.~2\textbf{fh} and Fig.~5. For each spectrum, the corresponding NC and the volume ratio $V/V_\V{Anderson}$ are shown on the right. Note that for the smallest NC (bottom black curve) no Coulomb peaks are observed, instead a large Coulomb gap and broad quantum well peaks are observed.}
	\end{center}
\end{figure}

\begin{figure}
	\begin{center}
		\includegraphics[width=16cm]{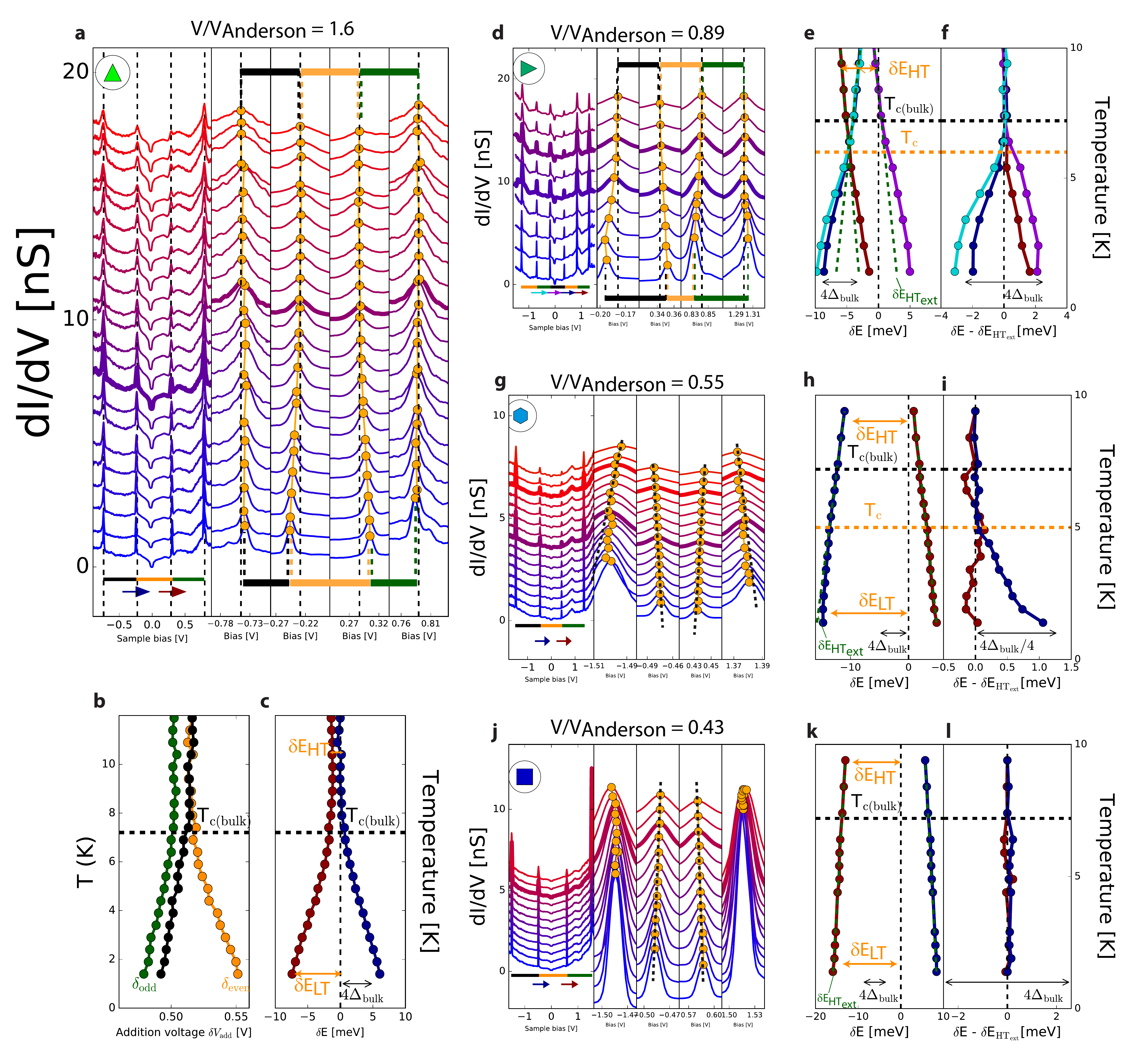}
		\caption{\label{Fig4} \textbf{Parity effect as function of temperature.} DC and addition energies for 4 NCs of decreasing volume, where $V/V_\V{Anderson}$ is indicated on top of the panels. (\textbf{a,d,g,j}) DC curves as function of temperature. The voltage separation between the Coulomb peaks, i.e. the addition voltage,  is indicated by the horizontal bars of different colors. In the same panels, zoom on the Coulomb peaks are shown where the maxima are indicated by orange dots. For panel \textbf{a}, the addition voltages are plotted as function of temperature on panel \textbf{b} with corresponding colors. The colored symbols (top left of panels) identify the corresponding data points in Fig.~2\textbf{fh} and Fig.~5.
		(\textbf{c,e,h,k}) Difference in addition energies between two charge configurations given by  $\delta E=\eta(\delta V_{\V{Head}}-\delta V_{\V{Tail}})$, where the head (tail) refers to the arrows shown in the corresponding panels. (\textbf{f,i,l}) Difference $\delta E - \delta E_{\V{HT}_{\V{ext}}}$ where the dash green line $\delta E_{\V{HT}_{\V{ext}}}$ is obtained from the extrapolation of $\delta E$ at high temperature. For panels \textbf{b,c,e,f,h,i,k,l}, the value $T_c(\V{bulk})$ is indicated as a black dash line. The extracted $T_{\V{c}}$ are shown as orange dash lines. A double-headed arrow provides the scale for the energy gap $4\Delta_{\V{bulk}}$ of bulk Pb.}
	\end{center}
\end{figure}

\begin{figure}
	\begin{center}
		\includegraphics[width=16cm]{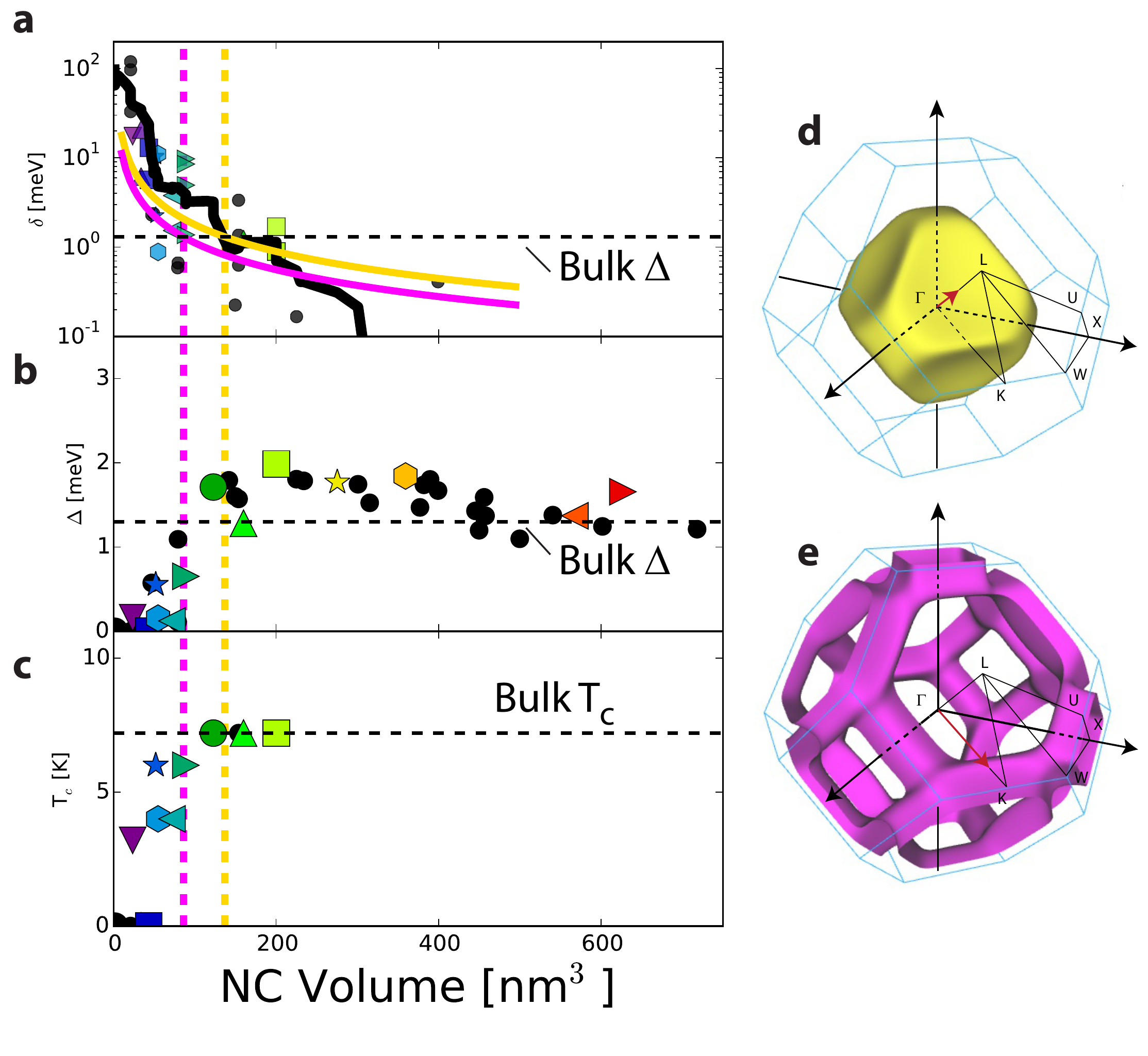}
		\caption{\label{Fig5} \textbf{Across the Anderson limit.} (\textbf{a}) Level spacing extracted from the addition energies measured above $T_{\V{c}}$. The experimental data (symbols) are highly scattered as a consequence of the random electronic level distribution. However, the average level spacing, shown by the smoothed black line, is of the order of magnitude of the calculated theoretical values shown as colored lines. The horizontal dash line indicates the bulk superconducting energy gap. (\textbf{b}) Superconducting gap $\Delta$ extracted from  the difference in addition energies between high and low temperature. The horizontal dash line indicates the bulk superconducting energy gap. (\textbf{c}) Transition temperature as function of NC volume. The horizontal dash line indicates the bulk transition temperature $T_\V{c}$=7.2 K. For all panels, the two vertical dash lines indicate the volumes where the level spacing reaches the superconducting energy gap at the wave vectors shown by red arrows on the two Fermi surfaces on the right. The colored symbols identify the corresponding DC curves in the other figures. For the black circles, the DCs are not shown. (\textbf{d}) Fermi surface (FS1) of the hole-type band of Pb. (\textbf{e}) Fermi surface (FS2) of the electron-type band of Pb.}
	\end{center}
\end{figure}

\subsection{References}
\bibliography{Bibliography}

\begin{addendum}
	\item[Acknowledgements] We acknowledge fruitful discussions with M. Aprili, C. Delerue and B. Grandidier. H.A., A.Z. acknowledge support from ANR grant "QUANTICON" 10-0409-01, ANR grant "CAMELEON" 09-BLAN-0388-01, China Scholarship Council and labex Matisse. DR and SP acknowledge C’NANO Ile-de-France, DIM NanoK, for the support of the Nanospecs project.
	
	\item[Author contributions] H.A. proposed the experiment. C.D., G.R., J.C.G designed the STM holder and prepared the InAs substrates. S.V., S.P., T.Z. and H.A. evaporated the lead nanocrystals on the cleaved InAs substrate and carried out the STM measurements with the help of D.R. H.A. analyzed the data and wrote the manuscript with the help of all authors.
	
	\item[Competing Interests] The authors declare that they have no
	competing financial interests.
	
	\item[Correspondence] Correspondence and requests for materials should be addressed to H.A.~(email: herve.aubin@espci.fr)

\end{addendum}

\clearpage

\end{document}

\section*{Supplementary Figures}

\begin{figure}[h!]
	\begin{center}
		\includegraphics[width=15cm]{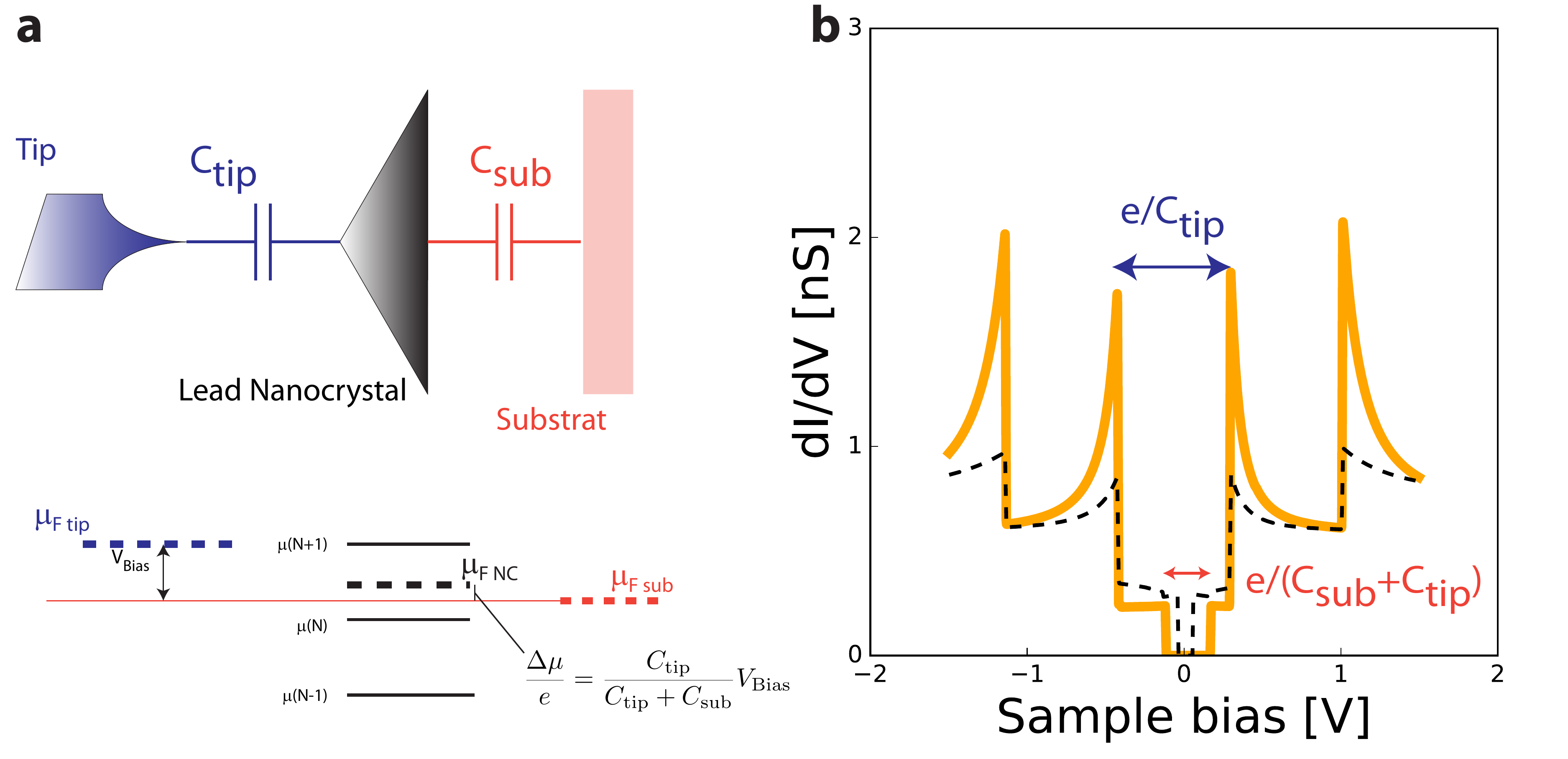}
		\caption{\label{FigS1} \textbf{Electrostatic model.} (\textbf{a}) Schematic of the double junction Tip-Nanocrystal-Substrate. (\textbf{b}) Simulation of the conductance spectrum using Hanna and Tinkham model\cite{Hanna1991} for two distinct values of the capacitance C$_{\V{sub}}$, shown by the continuous and dash lines. The voltage interval between the Coulomb peaks do not change with the capacitance C$_{\V{sub}}$, only the amplitude of the Coulomb gap at zero bias changes, as indicated by the double-headed arrow.}
	\end{center}
\end{figure}

\begin{figure}[h!]
	\begin{center}
		\includegraphics[width=10cm]{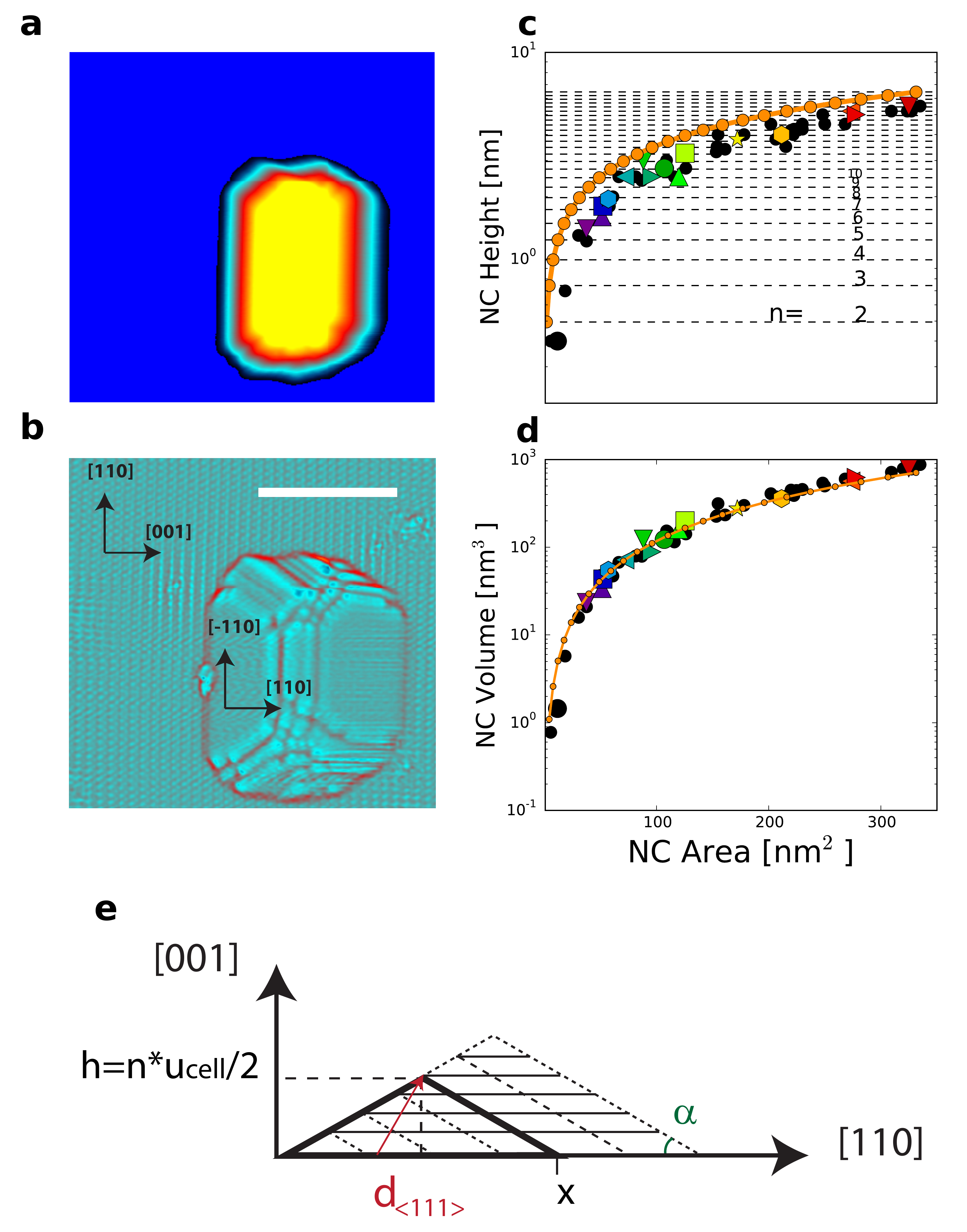}
		\caption{\label{FigS2} \textbf{Structural model of the nanocrystals.} (\textbf{a}) Flooded image used to determine the surface and volume of the nanocrystal shown panel \textbf{b}. (\textbf{b}) Laplacian image of a nanocrystal. The scale bar is 8 nm. (\textbf{c}) The symbols show the experimental nanocrystal height as function of nanocrystal area. The orange dots show the calculated height as function of calculated area for pyramidal nanocrystals of increasing height where n is the number of atomic Pb rows. The horizontal dash lines indicate the height corresponding to the number of atomic rows n. (\textbf{d}) The symbols show the experimental nanocrystal volume as function of nanocrystal area. The orange dots show the calculated volume as function of calculated area for pyramidal nanocrystals of increasing height. (\textbf{e}) is a schematic of the model used to calculate the height, the area and the volume of the pyramidal nanocrystal.}
	\end{center}
\end{figure}

\begin{figure}[h!]
	\begin{center}
		\includegraphics[width=12cm]{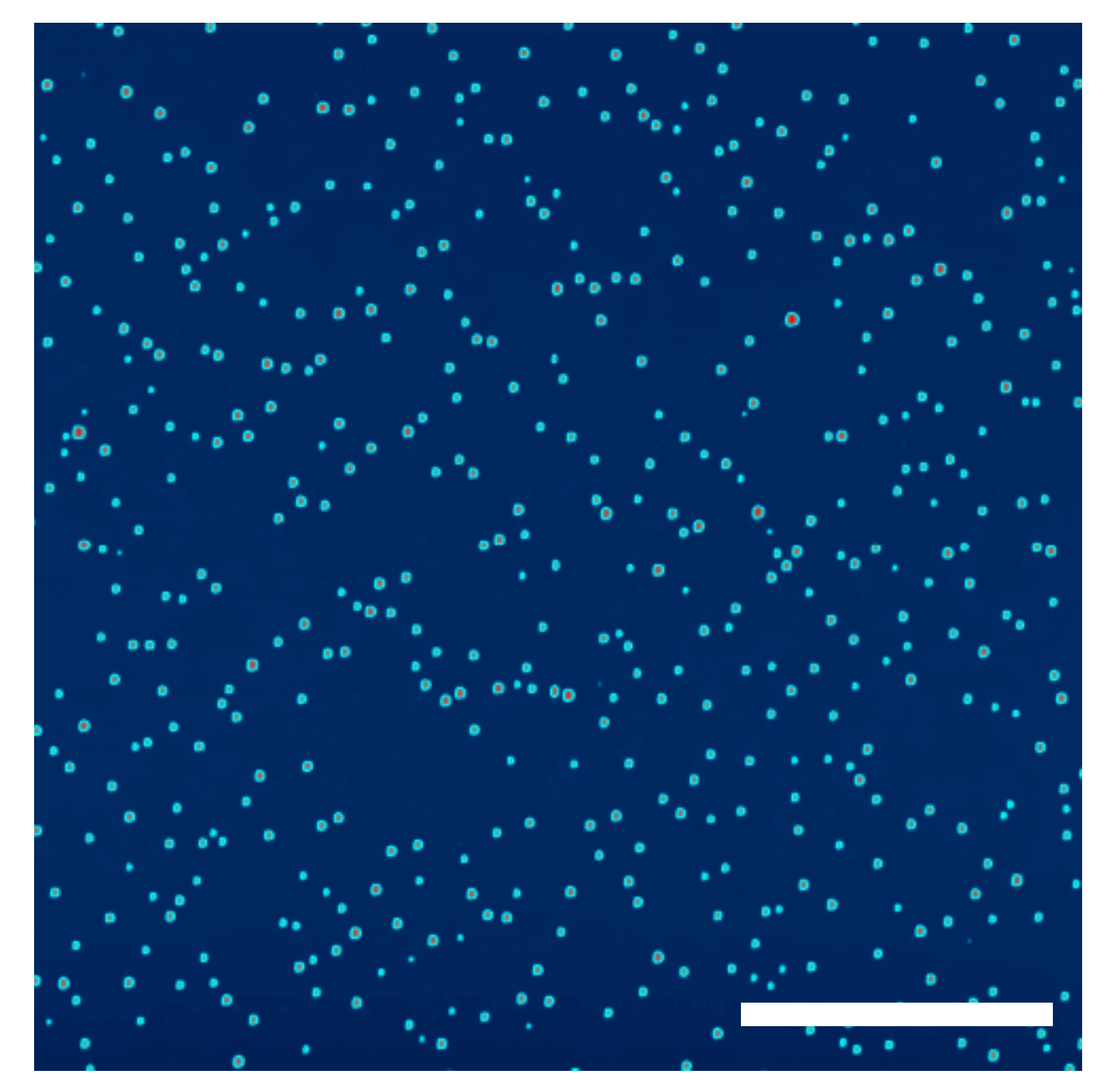}
		\caption{\label{FigS3} \textbf{Topographic image of sample B.} Large size 1~$\mu$m$\times$1~$\mu$m topographic STM image (1 V, 30 pA) of Pb nanocrystals grown on the (110) InAs surface of sample B. The scale bar is 300 nm.}
	\end{center}
\end{figure}

\begin{figure}[h!]
	\begin{center}
		\includegraphics[width=12cm]{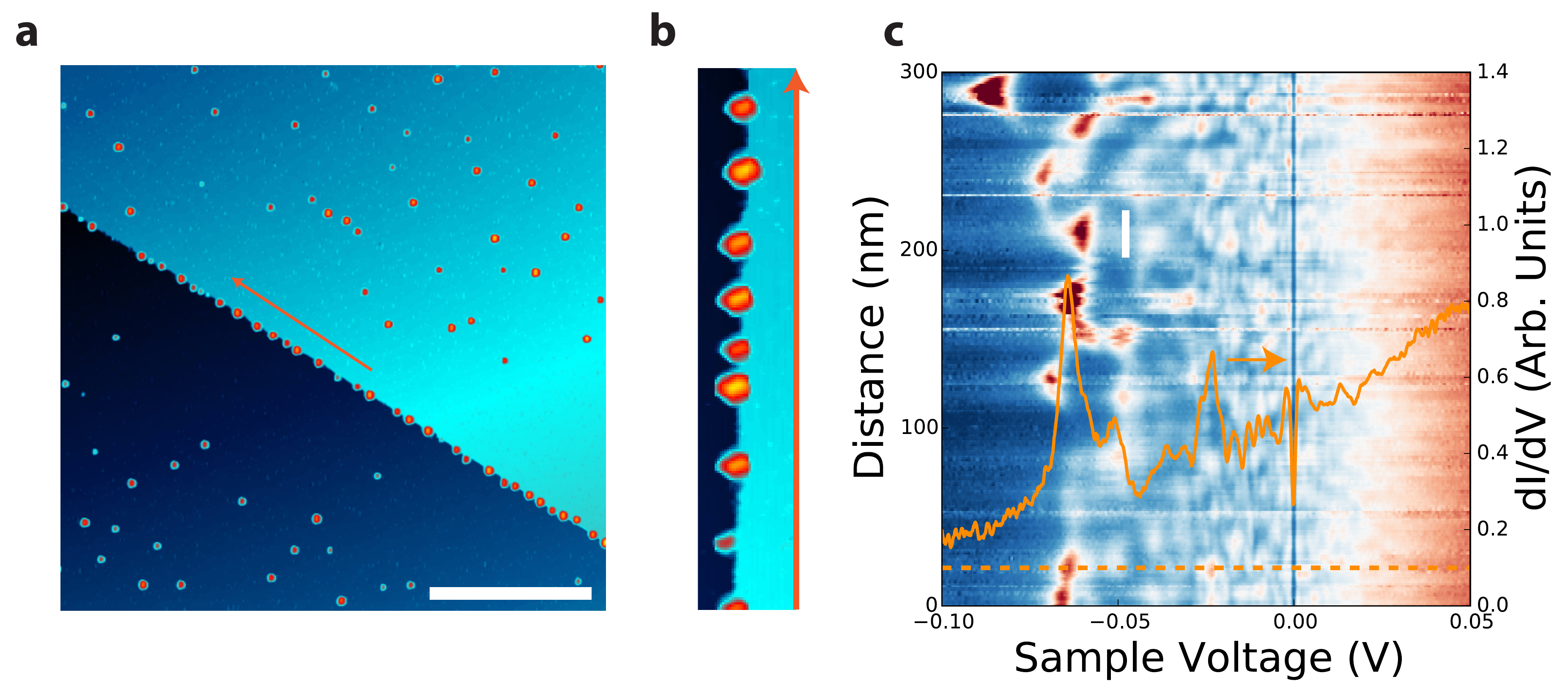}
		\caption{\label{FigS4} \textbf{Fluctuations of the energy of the tip-induced QDot levels.} (\textbf{a}) Large topographic image (1$\mu m \times 1 \mu m$) showing an atomic step edge against which the nanocrystals agglomerate. The scale bar is 300 nm. (\textbf{b}) Zoom on the area near the red arrow in panel \textbf{a} showing aligned nanocrystals along the atomic step edge.  (\textbf{c}) Conductance dI/dV map as function of sample voltage and distance measured along the red arrow shown in panel \textbf{b}. The orange line is the conductance curve extracted from the map at the location indicated by the horizontal dash line. The map shows that the tip-induced QDot levels are fluctuating in energy because of the presence of nearby nanocrystals. These fluctuations are long range ($>$ 30 nm). The scale bar is 30 nm. This length is of the order of the Fermi wavelength of the 2D electron gaz. These fluctuations are the consequence of the changing electrostatic environment due to the random distribution of the Pb nanocrystals and dopants.}
	\end{center}
\end{figure}

\begin{figure}[h!]
	\begin{center}
		\includegraphics[width=7cm]{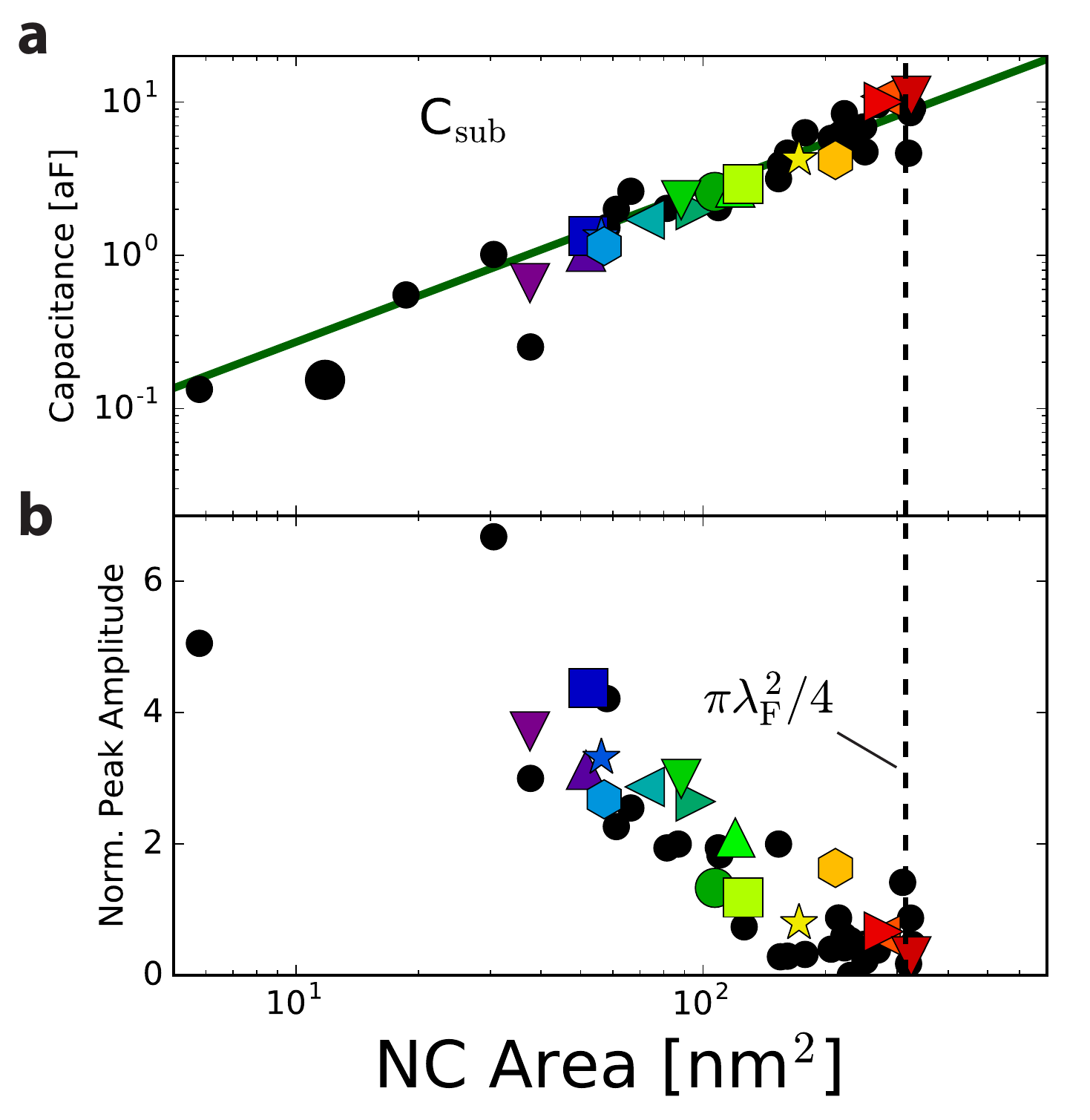}
		\caption{\label{FigS5} \textbf{Substrate-nanocrystal capacitance on a large range.} (\textbf{a}) The capacitance C$_{\V{sub}}$ extracted from the Coulomb gap at zero bias where the smallest nanocrystals have also been included.  (\textbf{b}) The normalized Coulomb peak amplitude decreases when the nanocrystal area approach $\pi\lambda_F^2/4$.}
	\end{center}
\end{figure}

\begin{figure}[h!]
	\begin{center}
		\includegraphics[width=12cm]{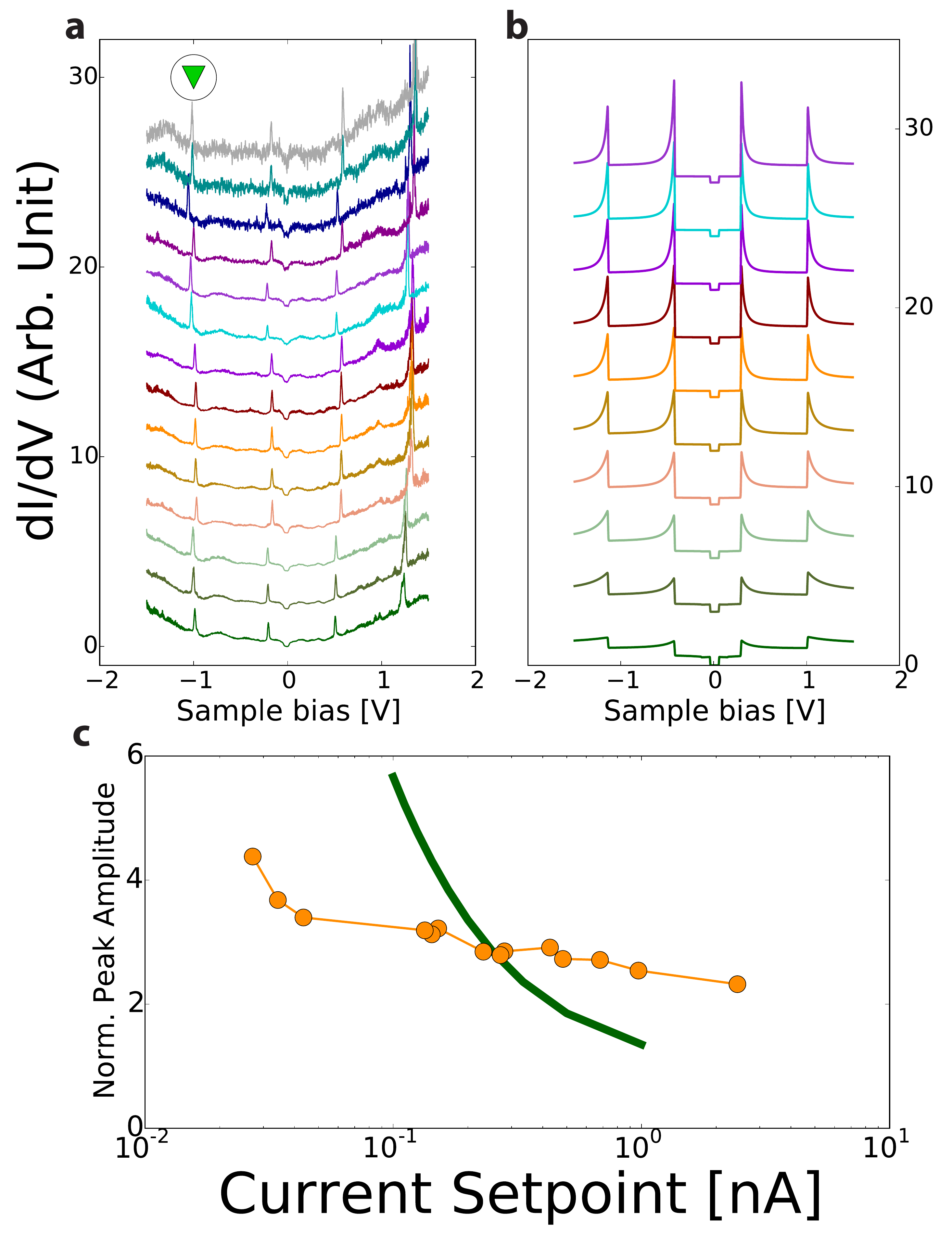}
		\caption{\label{FigS6} \textbf{Coulomb peaks as function of tip height.} (\textbf{a}) Normalized and shifted DC curves, measured at different setpoints from $I_{\rm set}$=5 nA (bottom) to $I_{\rm set}$=30 pA (top). (\textbf{b}) Corresponding theoretical DC curves obtained from the weak coupling model, Ref.~\cite{Hanna1991}. (\textbf{c}) Normalized peak amplitude measured experimentally (symbols) compared to the weak coupling model (line) of Hanna and Tinkham~\cite{Hanna1991}. This model can describe qualitatively the evolution of the Coulomb peak amplitude with the current setpoint, i.e. the peak amplitude is the largest for the highest tunnel junction resistance. However, this model is not sufficient to describe quantitatively the evolution of the peak amplitude.}
	\end{center}
\end{figure}

\begin{figure}[h!]
	\begin{center}
		\includegraphics[width=7cm]{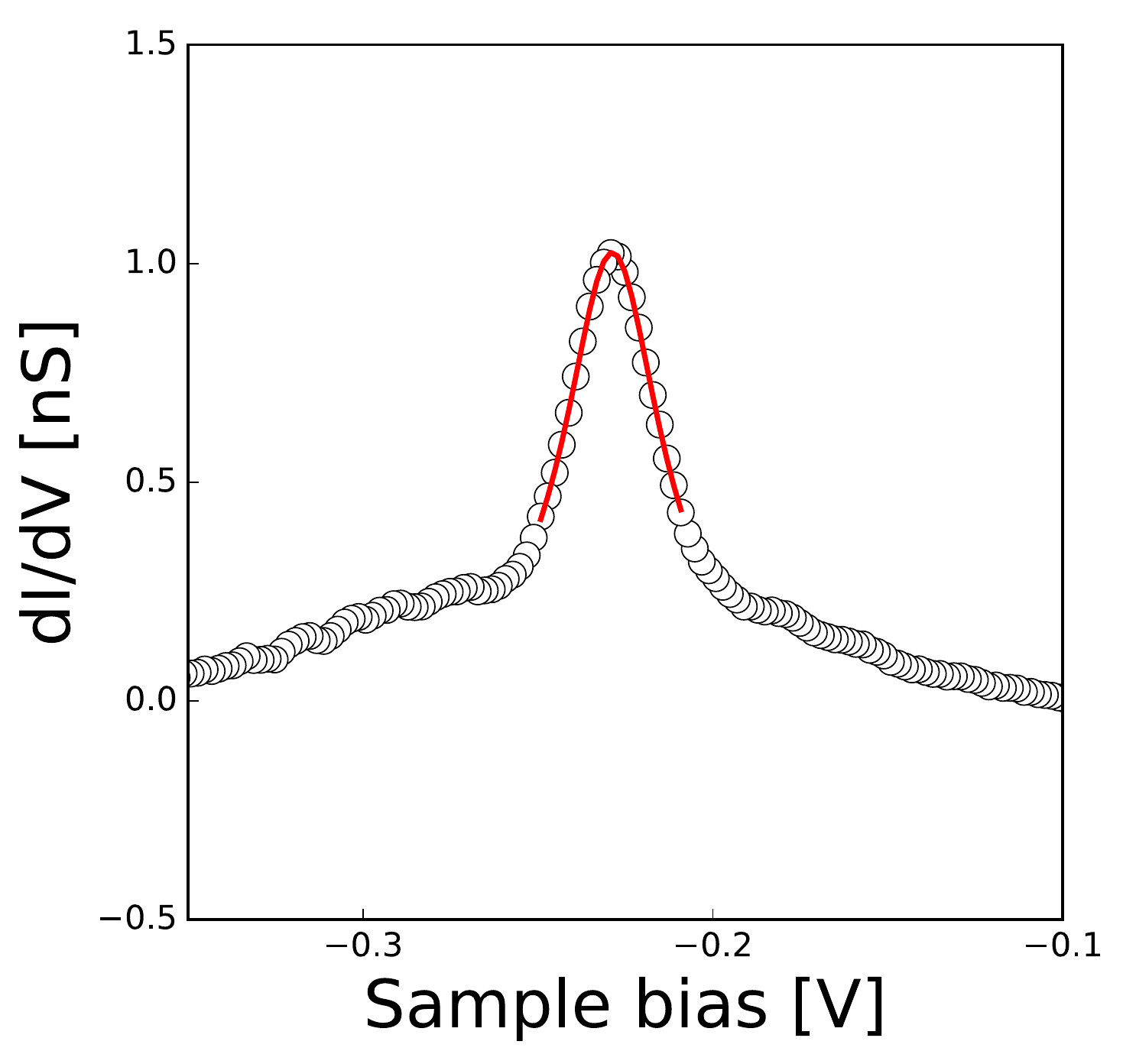}
		\caption{\label{FigS7} \textbf{Lorentz fit of the Coulomb peak.} Zoom on a single Coulomb peak. The voltage position of the Coulomb peak maxima is obtained through a fit with a Lorentz function.}
	\end{center}
\end{figure}

\begin{figure}[h!]
	\begin{center}
		\includegraphics[width=10 cm]{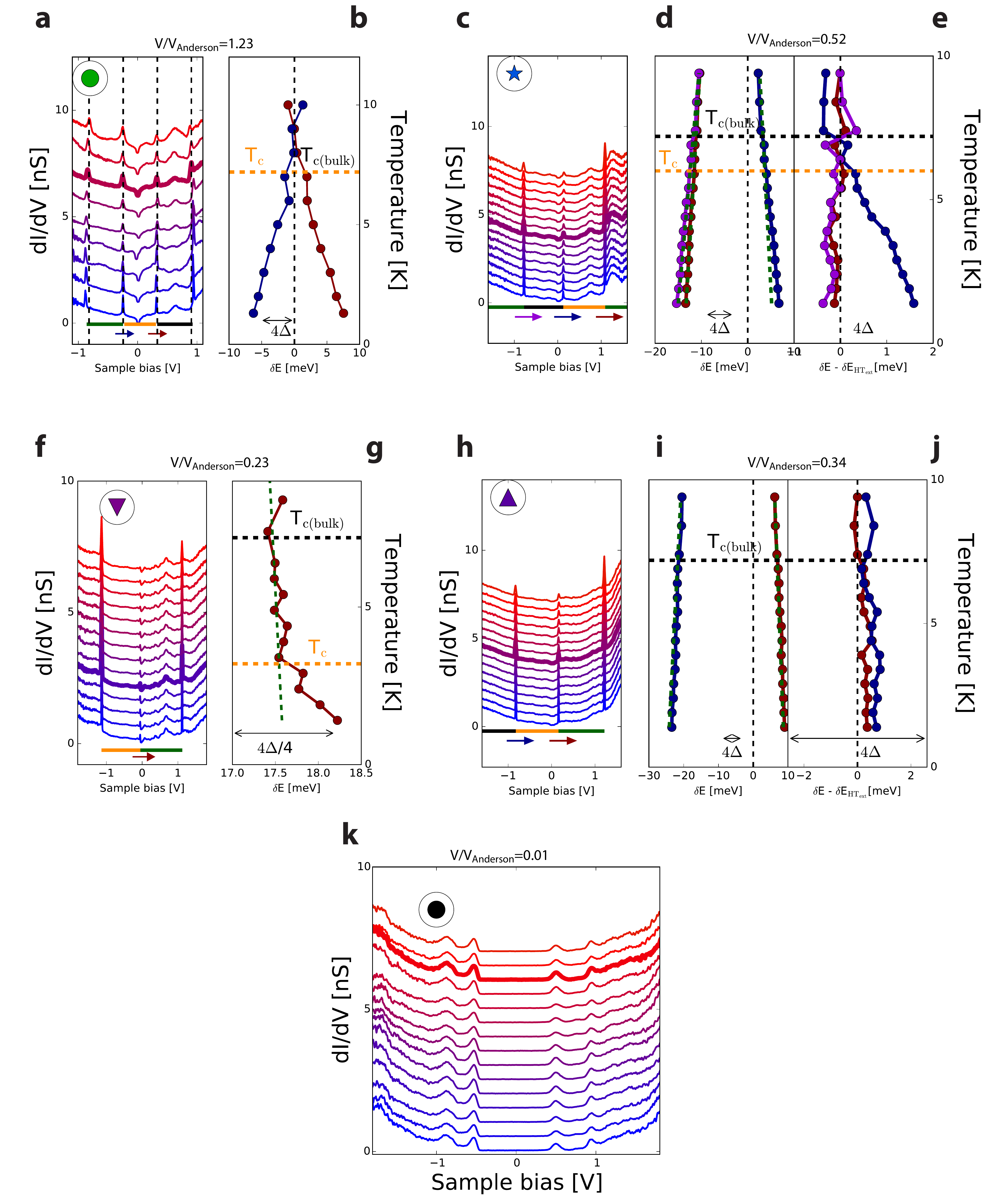}
		\caption{\label{FigS8} \textbf{Conductance spectrum of additional nanocrystals} (\textbf{a},\textbf{c},\textbf{f},\textbf{h},\textbf{k}) DC curves for 5 nanocrystals of decreasing volume where $V/V_{\rm Anderson}$ is indicated on top of the panels. (\textbf{b},\textbf{d},\textbf{g},\textbf{i}) Corresponding addition energies. The voltage separation between the Coulomb peaks is indicated by the horizontal bars of different colors. The difference in addition energies between two charge configurations is given by  $\delta E=(\delta V_{\rm Head}-\delta V_{\rm Tail})/\eta$, where the head (tail)  refers to the arrows shown in corresponding panels. (\textbf{e},\textbf{j}) Difference $\delta E - \delta E_{\V{HT}_{\V{ext}}}$ where $\delta E_{\V{HT}_{\V{ext}}}$ is obtained from the extrapolation of $\delta E$ at high temperature, shown as a dash green line. The $T_{\V{c(bulk)}}$ and energy gap $\Delta$ of bulk Pb are indicated in black. The extracted $T_{\V{c}}$ is shown as an orange dash line. (\textbf{k}) In this very small nanocrystal, $V\simeq 0.01 V_{\rm Anderson}$, no Coulomb peaks are observed, only the large Coulomb gap at zero bias and quantum well states are observed.}
	\end{center}
\end{figure}

\begin{figure}[h!]
	\begin{center}
		\includegraphics[width=10cm]{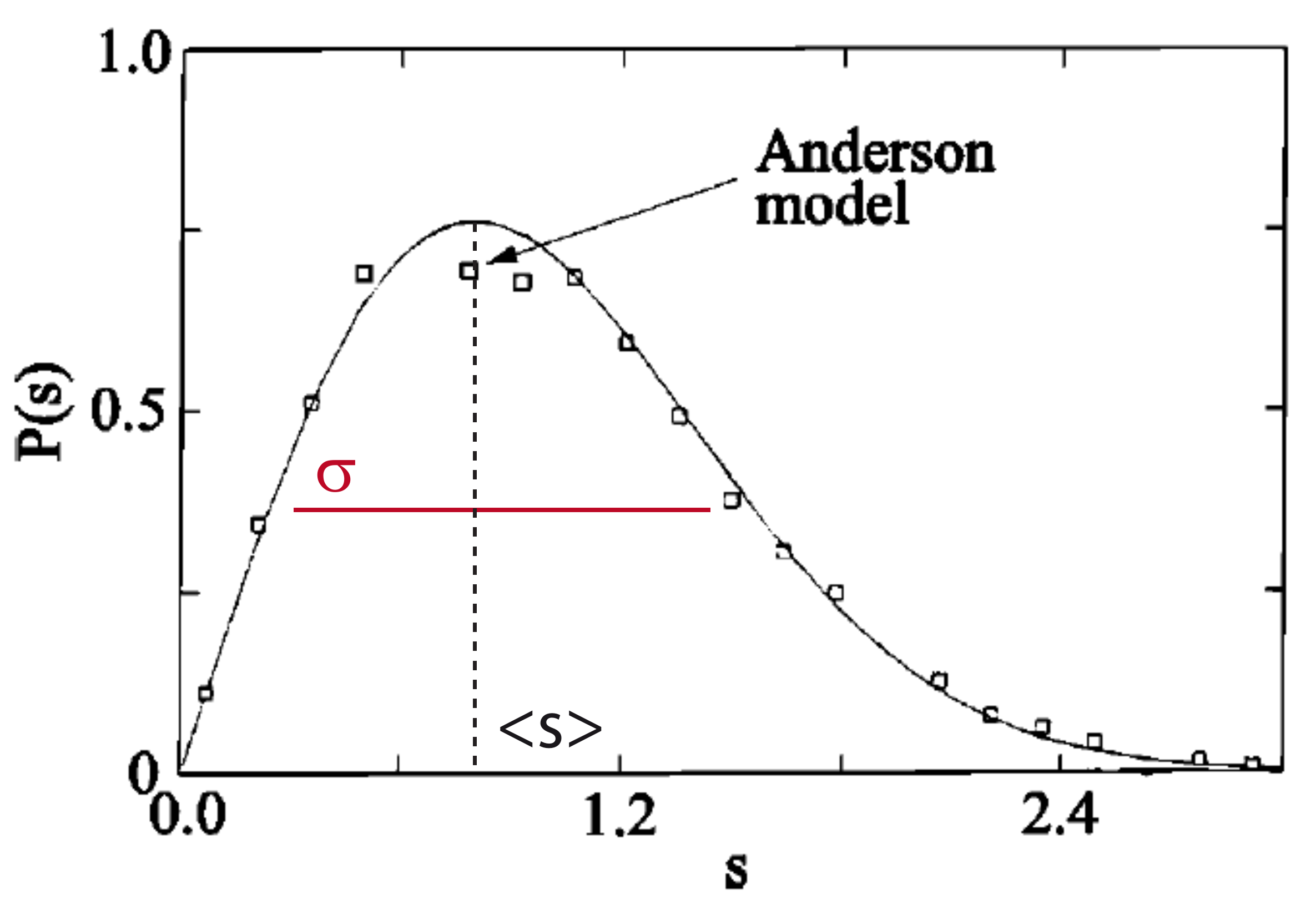}
		\caption{\label{FigS9} \textbf{Random Matrix Distribution.} Distribution $P(s)$ of electronic levels as function of level separation $s$ compared to the Anderson model, extracted from Ref.~\cite{Alhassid2000}. The width $\sigma$ of the distribution is of the order of the level spacing.}
	\end{center}
\end{figure}

\clearpage

\section*{Supplementary Notes}

\subsection{Supplementary Note 1. Structural model of the nanocrystals.}

The surface and the volume of the nanocrystals are obtained by a flooding method, Supplementary Fig. 2, as found in most STM images analysis softwares. The height of the nanocrystals and their volume are plotted Supplementary Fig.~2c and Supplementary Fig.~2d, respectively, as function of the nanocrystals area. As described in the main text, the nanocrystals facets are mostly oriented along the [111] direction. Thus, the shape of the nanocrystals is mostly pyramidal as sketched Supplementary Fig.~2e, where the [001] direction of Pb is oriented perpendicular to the substrate.
For this geometry, the height of the nanocrystal is given by $h=n\times u_{\V{cell}}/2$ where $u_{\V{cell}}$=0.495 nm is the length of the unit cell of Pb; $u_{\V{cell}}/2$ is the distance between atomic rows along the [001] direction of Pb, which has a Face Centered Cubic (FCC) structure, and n is the number of atomic rows. Assuming that the nanocrystal is a perfect pyramid, its area can be calculated from the relation $s=x^2$ with $x=2\times h/\tan{(\alpha)}$, where the definition of $\alpha$ is given in Supplementary Fig.~2e. Furthermore, the volume of the pyramid can be calculated from $v=s\times h/3$. Plotting the calculated height as function of calculated area together with the experimental data, Supplementary Fig.~2d shows that this model can describe nicely the evolution of the nanocrystal height with the area of its base. Deviations from this model are expected since the nanocrystals are not perfect pyramids, their top are usually truncated. It can be noticed that the smallest nanocrystals measured are only two atomic rows high, i.e. one unit cell high. The spectra of these nanocrystals, shown Figure 3 (bottom curve) and Supplementary Fig. 8k, do not present any Coulomb peaks but only broad peaks due to the formation of quantum well states.

\subsection{Supplementary Note 2. Comparison with the results of Bose et al.\cite{Bose2010}}

In our work the nanoparticles are only weakly coupled to the conducting substrate as demonstrated by the presence of the Coulomb gap and the sharp Coulomb peaks. In the work of Bose et al.\cite{Bose2010}, no Coulomb gap or Coulomb peaks are observed, indicating that their nanoparticles are strongly coupled to the substrate. This is a fundamental difference between the two systems and has deep consequences on the evolution of the superconducting characteristics ($T_c$ and $\Delta$) with the size of the nanoparticle.

In our data, both quantities, $T_c$ and $\Delta$, do not change with the volume from 800 nm$^3$ down to the Anderson volume 100 nm$^3$. At the Anderson volume, both quantities go to zero very sharply. This suppression of superconductivity results from the suppression of pairing when the energy interval between two electronic levels becomes larger than the superconducting gap. This is the physical origin of the Anderson criterion, i.e. only levels within the superconducting gap energy form  Cooper pairs.

In contrast, in the data of Bose et al., the superconducting gap value starts to decrease below a nanoparticle height about 10 nm, which corresponds to a volume of 2000 nm$^3$ for a semi-spherical shape, as indicated in their paper. Thus, in their system, the amplitude of the superconducting gap starts decreasing at a volume about 20 times larger than the Anderson volume. The authors attribute this reduction to quantum fluctuations of the superconducting order parameter. We could add, that, generally, the superconducting gap in strongly coupled normal-superconducting structures is expected to be smaller than the pure bulk superconductor because of the inverse proximity effect from the normal region onto the superconducting region.

The paper by Bose et al. also reports on the observation of the shell effect in Sn nanoparticles and the absence of this shell effect in the Pb nanoparticles. Within our experimental resolution, we did not observe the shell effect in the Pb nanoparticles either.

\bibliographystyle{naturemag}
\bibliography{Bibliography}